# On the standard quantum Brownian equation and an associated class of non-autonomous master equations

Allan Tameshtit
Email: allan.tameshtit@utoronto.ca

It is shown that the standard quantum Brownian equation (QBE) can violate positivity not only past the thermal correlation time, but at arbitrarily long times at high system frequencies. In an effort to improve the standard QBE, exact operator solutions are provided for a class of non-autonomous master equations. These exact solutions are used to derive sufficient positivity conditions for the coefficients of the master equations.

## Introduction

When diffusion and dissipation are to be included in non-relativistic dynamics, perhaps the most widely used equation is the Kramers equation [1] in phase space, first investigated by workers in the first half of the twentieth century [2]. The extension of this equation to quantum mechanics for a general potential was derived by Caldeira and Leggett [3]. For a harmonic oscillator, the equation takes the form of the standard quantum Brownian equation (QBE) [4,5]:

$$\frac{d\rho}{dt} = \frac{1}{\hbar i}[H,\rho] + \frac{\Gamma}{2\hbar i}\{p,\rho,q\} - \frac{\Gamma}{2\hbar i}\{q,\rho,p\} - \frac{2\Gamma m kT}{\hbar^2}\{q,\rho,q\}, \quad (1)$$

where $\rho$ is the density operator, $\Gamma$ is a positive coupling constant, $\{A,\rho,B\} \equiv BA^\dagger\rho + \rho BA^\dagger - 2A^\dagger\rho B$ [6] and

$$H = \frac{p^2}{2m} + \frac{\Gamma}{2}(qp+pq) + \frac{1}{2}m\omega^2 q^2. \quad (2)$$

Eq. (1) can be derived by a variety of methods by invoking several approximations that usually involve high temperatures and sufficiently long times. Pushing the equation beyond these limits introduces anomalies that can include the emergence of non-positive density operators [5,7,8,9]. Unfortunately, with several approximations in play, it is not a simple matter to quantitatively characterize the regime in which positivity, at least, is preserved by Eq. (1). Comments in Ref. [10] in connection with a Brownian free particle might prompt one to conclude that positivity is ensured if two conditions are fulfilled:

$$\frac{kT}{\hbar\Gamma} \gg 1 \quad (3)$$

and

$$t \gg \frac{\hbar}{kT}. \quad (4)$$

The first, a high temperature requirement, is deemed to be necessary to justify a cumulant expansion that can lead to the irreversible terms of Eq. (1). The second condition recognizes that positivity failure can occur at times that are short compared to the thermal correlation time $\hbar/kT$. However, below we show that conditions (3) and (4) are not a cure for the positivity problem for the harmonic oscillator, and, surprisingly, that non-positivity can arise at arbitrarily long times even if these conditions hold, albeit at large oscillator frequencies and for certain initial conditions [11]. Along the way, we present a necessary and sufficient condition for Eq. (1) to preserve positivity.

Instead of pushing Eq. (1) to its limits to try to expand its scope, another approach that has been adopted to try to remedy positivity failure involves adding a term proportional to $\{p,\cdot,p\}$ [13,14]. This addition is usually made axiomatically, but an attempt has also been made to derive such a term [15]. In the next section, we take a critical look at this approach.

Exact non-autonomous master equations are known for quadratic systems [5,16,17]. To increase the scope of the approximate Eq. (1), one can attempt to modify these exact master equations by introducing approximations that are less drastic than those leading to Eq. (1). Although one may question the need to modify an exact result, the motivation for doing so arises because the coefficients of the exact master equations are quite unmanageable. For these equations to be of any practical use and to be able to gain insights buried in the exact expressions, approximations are usually necessary.

And therein lies a problem: exact equations preserve positivity *ipso facto*. As soon as any simplifying approximations are introduced, the preservation of positivity is not guaranteed. To provide guidelines for acceptable approximations, below we deduce sufficient conditions for the coefficients of a class of generally non-autonomous master equations that ensure positivity. Positivity conditions in non-autonomous/non-Markovian master equations have also been examined in [18]. Our approach for formulating such sufficient conditions requires us to first find exact solutions for the class of master equations, which solutions we provide below in operator form. But first, let us turn back to Eq. (1) and examine more closely the two putative remedies therefor.

*Addition of $\{p,\cdot,p\}$ term*

The desire to add a term proportional to $\{p,\cdot,p\}$ to Eq. (1) stems from the work in [19]. There, Lindblad examined bounded operators (it has often been assumed that the results there apply to unbounded operators as well) and proved for evolution obeying the semi-group property that a necessary and sufficient condition for a master equation to yield a completely positive density operator is that the equation have the form

$$\frac{d\rho}{dt} = \frac{1}{\hbar i}[H,\rho(t)] - \sum_{\alpha}\{C_{\alpha},\rho(t),C_{\alpha}\} \qquad (5)$$

where the $C_\alpha$ are time-independent operators. As the machinery in Lindblad's work is involved, let us examine a heuristic argument [20] that shows how the form of Eq. (5) arises.

In the absence of any coupling between the system of interest and environment, suppose $\rho(t)$ were governed by $d\rho(t)/dt = [H, \rho(t)]/\hbar i$, where $H$ is time-independent. With the coupling turned on, this last equation is modified, but assume that the modified propagator satisfies the semi-group property, $\mathcal{U}(t_1 + t_2) = \mathcal{U}(t_1)\mathcal{U}(t_2)$ for $t_1, t_2 \geq 0$, and that the evolution of $\rho(t)$ is completely positive. This means that there exist operators $W_\alpha(t)$ such that $\rho(t) = \mathcal{U}(t)\rho(0) = \sum_\alpha W_\alpha^\dagger(t)\rho(0)W_\alpha(t)$ with $\sum_\alpha W_\alpha(t)W_\alpha^\dagger(t) = 1$. Define the interaction picture as $\rho^I(t) = U^{-1}(t)\rho(t)U(t)$ where $dU(t)/dt = HU(t)/\hbar i$. We compute

$$\frac{d\rho^I(t)}{dt} = \lim_{\Delta t \downarrow 0} \frac{\rho^I(t + \Delta t) - \rho^I(t)}{\Delta t} \tag{6}$$

$$= \lim_{\Delta t \downarrow 0} \sum_\alpha [U^{-1}(t + \Delta t)W_\alpha^\dagger(\Delta t)\rho(t)W_\alpha(\Delta t)U(t + \Delta t)^\dagger$$

$$- \frac{1}{2}U^{-1}(t)W_\alpha(\Delta t)W_\alpha^\dagger(\Delta t)\rho(t)U(t) - \frac{1}{2}U^{-1}(t)\rho(t)W_\alpha(\Delta t)W_\alpha^\dagger(\Delta t)U(t)]/\Delta t \tag{7}$$

Introducing the decomposition $W_\alpha(t) = A_\alpha(t) + c_\alpha(t)I$, we obtain

$$\frac{d\rho^I(t)}{dt} =$$

$$\sum_\alpha \lim_{\Delta t \downarrow 0} \frac{1}{2\Delta t} U^{-1}(t)(-\{A_\alpha(\Delta t), \rho(t), A_\alpha(\Delta t)\} + [c_\alpha(\Delta t)A_\alpha^\dagger(\Delta t) - c_\alpha^*(\Delta t)A_\alpha(\Delta t), \rho(t)])U(t)$$

$$\tag{8}$$

Now assuming

$$\lim_{\Delta t \downarrow 0} \frac{1}{2\Delta t} \{A_\alpha(\Delta t), \rho(t), A(\Delta t)\} = \{C_\alpha, \rho(t), C_\alpha\} \tag{9}$$

and

$$\sum_\alpha \lim_{\Delta t \downarrow 0} \frac{1}{2\Delta t} [c_\alpha(\Delta t)A_\alpha^\dagger(\Delta t) - c_\alpha^*(\Delta t)A_\alpha(\Delta t), \rho(t)] = \frac{1}{\hbar i}[H', \rho(t)] \tag{10}$$

we get

$$\frac{d\rho^I(t)}{dt} = \frac{1}{\hbar i} U^{-1}(t)[H', \rho(t)]U(t) - \sum_\alpha U^{-1}(t)\{C_\alpha, \rho(t), C_\alpha\}U(t) \tag{11}$$

Reverting back to the Schrodinger picture, we obtain

$$\frac{d\rho(t)}{dt} = \frac{1}{\hbar i}[H + H', \rho(t)] - \sum_\alpha \{C_\alpha, \rho(t), C_\alpha\}, \tag{12}$$

which has the same form as Eq. (5). If we were to compare the standard QBE to this last equation for the special case $C_\alpha = a_\alpha q + b_\alpha p$, we would notice that the standard QBE is missing a term proportional to $\{p, \cdot, p\}$. This has motivated many workers to add such a term, although some authors have remarked that for Brownian motion a $\{p, \cdot, p\}$ term is difficult to justify physically [10]. In Ref. [15], an attempt was made to derive such a term. Unfortunately, the derivation requires that some terms be omitted that are no smaller than those appearing in the final equation [21].

For the most part, a term proportional to $\{p,\cdot,p\}$ has arisen in the quantum Brownian literature by fiat. For example, in Ref. [22], the term $-\gamma\{p,\cdot,p\}/8mkT$, where $\gamma$ is a characteristic damping rate, was added to Eq. (1) to ensure that positivity is preserved. However, one drawback of this procedure is that it is not unique. For example, when added to Eq. (1), any term $-b\{p,\cdot,p\}$ where $b \geq \Gamma/8mkT$ ensures positivity preservation [23].

In any case, even with a $\{p,\cdot,p\}$ term, an autonomous equation having the semi-group property and of Lindblad form (i.e. of the form of Eq. (5)) cannot be valid at short times when such an equation derives from a total Hamiltonian with factorized coupling and initial condition [21]. To see this, compute $\frac{d}{dt}Tr\rho^2$ at the initial time. The answer is zero. Physically this means that instantaneously the system behaves reversibly when initially pure, the bath interactions having not yet had a dissipative effect. On the other hand, at the initial time and with initial normalized condition $|\psi\rangle\langle\psi|$, Eq. (5) yields $\frac{d}{dt}Tr\rho^2 = -2\sum_\alpha \langle\psi|\{C_\alpha,|\psi\rangle\langle\psi|,C_\alpha\}|\psi\rangle$, which is typically negative. Recognizing discrepancies predicted by Eq. (1) in the inner limit, workers have suggested that Eq. (1) is only valid at longer times under certain conditions. We next examine this statement more closely.

*Non-positivity at arbitrarily long times for the standard QBE*

To determine *if* Eq. (1) violates positivity, it is sufficient to examine expectation values at $t = 0$, from which it can be concluded that it does [9]. To determine *when* Eq. (1) violates positivity, it is necessary to examine behavior at arbitrary times. In Ref. [24], a version of Eq. (1), in which $kT$ is replaced by $\hbar\omega_0\left[\left(e^{\hbar\omega/kT}-1\right)^{-1}+1/2\right]$, was used to analyze the temporal behavior of positivity for a particular class of initial conditions (squeezed states). In this section, we relax this last constraint and analyze in a different manner the temporal behavior of positivity for any initial condition that evolves according to Eq. (1).

In the interaction picture, where $\rho^I(t) = e^{it[H,\cdot]/\hbar}\rho(t)$, and using the techniques expounded in Ref. [25], we find the following solution of Eq. (1):

$$\rho^I(t) = \exp[\Delta(t)\{p,\cdot,p\}]\exp[-\Gamma t\{\hat{B}^\dagger,\cdot,\hat{B}^\dagger\}]\rho(0) \tag{13}$$

where

$$\Delta(t) = \left(\frac{1}{\tilde{\eta}\hbar}\frac{kT}{\hbar\omega}\right)^2 \frac{1}{l_1(t)}\left(\sin^2\!\left(\tilde{\eta}u\right) - r^2\sinh^2 u\right) \tag{14}$$

with $u = \Gamma t$, $\tilde{\eta} = \sqrt{\omega^2/\Gamma^2 - 1}$, $r = \tilde{\eta}\sqrt{1 - \left(\frac{\hbar\omega}{2kT}\right)^2}$ (both $\tilde{\eta}$ and $r$ assumed real), and

$$l_1(t) = \frac{mkTe^{2\Gamma t}}{2\hbar^2}\left[1 - e^{-2\Gamma t} + \left(1 - \cos\!\left(2\tilde{\eta}\Gamma t\right)\right)/\tilde{\eta}^2 + \sin\!\left(2\tilde{\eta}\Gamma t\right)/\tilde{\eta}\right],$$

and where the generalized lowering operator (satisfying $[\hat{B}, \hat{B}^\dagger] = 1$) may be written as

$$\hat{B}(t) = \left(\frac{2l_1(t)}{e^{2\Gamma t} - 1}\right)^{1/2} q + \left(\frac{e^{2\Gamma t} - 1}{2l_1(t)}\right)^{1/2}\left(\frac{2l_3(t)}{e^{2\Gamma t} - 1} + \frac{i}{2\hbar}\right)p \tag{15}$$

with $l_3(t) = \frac{kTe^{2\Gamma t}}{2\hbar^2\Gamma\tilde{\eta}^2}\left[1 - \cos\!\left(2\tilde{\eta}\Gamma t\right)\right]$, although other expressions for $\hat{B}(t)$ could have been chosen (see Appendix 2 for a related discussion). The sign of $\Delta(t)$ is determined by $\sin^2\!\left(\tilde{\eta}u\right) - r^2\sinh^2 u$, since $l_1(t) > 0$ for $t > 0$.

Paying heed to non-commuting operators, suppose we were to combine the two exponents of Eq. (13) into one: $\rho'(t) = e^{L(t)}\rho(0)$. Then the theorem in Appendix 1 and the corollary in Ref. [26] (see also [25]) allow us to paint the following picture. At times when $\sin^2\!\left(\tilde{\eta}u\right) - r^2\sinh^2 u < 0$, signifying that there are more than enough fluctuations to preserve positivity at such times, $L(t)$ can be written as $-\sum_{n=1}^{2}\{a_n q + b_n p, \cdot, a_n q + b_n p\}$, where the $a_n$ and $b_n$ are time-dependent, generally complex numbers, and positivity is preserved. When $\sin^2\!\left(\tilde{\eta}u\right) - r^2\sinh^2 u$ is zero, say at time $t'$, $L(t')$ is given by

$-\Gamma t'\{\hat{B}^\dagger(t'), \cdot, \hat{B}^\dagger(t')\}$, and there are just enough fluctuations to preserve positivity. In this case, it is also of interest to note that there exist initial states that evolve after a time $t'$ to pure states [27]. In particular, suppose first we take for an initial state $|\beta(t')\rangle\langle\beta(t')|$ where $|\beta(t')\rangle$ is an eigenvector of $\hat{B}(t')$ with eigenvalue $\beta(t')$. Such states $|\beta\rangle$, dubbed two-photon coherent states in the quantum optics literature, have been studied in [28]. Next, consider the following identities, which are suggested from results in Zel'dovitch *et al*. [29] and which after the fact can be proved by introducing a parameter in the exponential on the left-hand side and showing that both sides of the resultant identities obey the same differential equation:

$$\exp[-r\{A^\dagger, \cdot, A^\dagger\}] = e^{-rA^\dagger A}[\exp[(1 - e^{-2r})A \cdot A^\dagger]\rho]e^{-rA^\dagger A} \tag{16}$$

and
$$\exp[-r\{A, \cdot, A^\dagger\}] = e^{-rAA^\dagger}[\exp[(1-e^{-2r})A^\dagger \cdot A]\rho]e^{-rAA^\dagger} \tag{17}$$

where $A$ is any operator satisfying $[A, A^\dagger]=1$ and $r \geq 0$ is a c-function [30]. Using relation (16), we get

$\rho^I(t') = \exp[-\Gamma t'\{\hat{B}^\dagger(t'), \cdot, \hat{B}^\dagger(t')\}]|\beta(t')\rangle\langle\beta(t')| = |\beta(t')e^{-\Gamma t'}\rangle\langle\beta(t')e^{-\Gamma t'}|$. Finally, if $\sin^2\left(\tilde{\eta}u\right) - r^2\sinh^2 u$ is positive at a specific time, then

$L(t) = -R_1(t)\{q, \cdot, q\} + R_3(t)\{q, \cdot, p\} + R_3^*(t)\{p, \cdot, q\} - R_2(t)\{p, \cdot, p\}$, where $R_1$ and $R_2$ are non-negative parameters, and $R_3$ a complex parameter with $R_1 R_2 - |R_3|^2$ being less than zero, signifying that there are not enough fluctuations for positivity to hold at that specific time for all initial states. To summarize, for evolution governed by Eq. (1), the density operator is positive at a particular time for all initial states if and only if $\sin^2\left(\tilde{\eta}u\right) - r^2\sinh^2 u \leq 0$ at that particular time.

These results make clear under what conditions non-positivity can linger. As an example, let $r \to 0^+$ (by letting $\dfrac{\hbar\omega}{2kT}$ approach unity from below), and fix $\tilde{\eta}$ to satisfy condition (3). Then the maximum $u$ at which $\sin^2\left(\tilde{\eta}u\right) - r^2\sinh^2 u$ is positive grows without bound. And this trick can be performed while satisfying conditions (3) and (4), which demonstrates that these conditions alone cannot generally ensure positivity.

The foregoing analysis should not be taken to imply that there do not exist regions of parameter space that ensure positivity for Eq. (1) at sufficiently long times (regions ensuring positivity are precisely those for which $\sin^2\left(\tilde{\eta}u\right) - r^2\sinh^2 u \leq 0$). Rather, we have just seen that the particular conditions (3) and (4) are not sufficient.

To see how one can improve the standard QBE to make it more universally valid, we next present some exact operator solutions for a class of non-autonomous master equations.

*Exact solutions of a class of master equations*

Exact solutions of autonomous harmonic oscillator master equations have been found using operator [31] and path integral [32] techniques. However, when starting from a total Hamiltonian consisting of a harmonic oscillator coupled to a reservoir, the exact master equations derived therefrom turn out to be non-autonomous and typically of the form

$$\frac{d}{dt}\rho(t)=\frac{1}{\hbar i}[H_s(t),\rho(t)]-k_1(t)\{q,\rho(t),q\}-k_2(t)\{p,\rho(t),p\}+k_3(t)\{p,\rho(t),q\}+k_4(t)\{q,\rho(t),p\}$$
(18)

where $H_s(t) \equiv b_{11}(t)q^2 + b_{12}(t)(qp+pq) + b_{22}(t)p^2$, with $b_{11}, b_{12}, b_{22}, k_1$ and $k_2$ being real continuous functions, and with $k_3$ and $k_4$ being complex continuous functions such that $k_3^* = k_4$ [5,16]. As a notable example, Eq. (1) derives from such equations after certain limiting approximations [5].

A non-autononomous model, in which the Hamiltonian is time-dependent but the irreversible terms are the familiar time-independent ones from the quantum optical master equation, was investigated in Ref. [33] using operator techniques similar to the ones used below. However, we work in the usual space of density operators, instead of a "super-Hilbert space" of density operator kets $|\rho\rangle$ considered in [33]. In this vein, a "superoperatorial" approach was used to treat certain non-Markovian master equations in Ref. [34]. Brownian evolution has also been previously examined using the Wei-Norman method [25] for solvable Lie algebras that we use below. In some recent work [35], this method has been used to make some general observations for certain non-autonomous master equations and to analyze a two-level spin system.

In this section, we turn our attention to Eq. (18), and first remark that the desire to simplify the coefficients therein (the $k$'s in Eq. (18)) is understandable because these coefficients are typically unwieldy. In addition, by simplifying the coefficients, the important underlying physics becomes transparent. What would therefore be desirable is to approximate the $k$'s while ensuring that solutions of Eq. (18) remain positive. We will proceed to characterize the types of approximations that preserve positivity, but first we need to find exact solutions for Eq. (18).

The operators in the master equation (18) form a closed algebra in view of the following table of commutators [25,36]:

|  | $L_s$ | $\{q,\cdot,q\}$ | $\{p,\cdot,p\}$ | $\{p,\cdot,q\}$ | $\{q,\cdot,p\}$ |
|---|---|---|---|---|---|
| $L_s$ | 0 | $-4b_{12}\{q,\cdot,q\}$ $-2b_{22}(\{p,\cdot,q\}+\{q,\cdot p\})$ | $4b_{12}\{p,\cdot,p\}$ $+2b_{11}(\{p,\cdot,q\}+\{q,\cdot p\})$ | $2b_{11}\{q,\cdot,q\}$ $-2b_{22}\{p,\cdot,p\}$ | $2b_{11}\{q,\cdot,q\}$ $-2b_{22}\{p,\cdot,p\}$ |
| $\{q,\cdot,q\}$ |  | 0 | 0 | $2\hbar i\{q,\cdot,q\}$ | $-2\hbar i\{q,\cdot,q\}$ |
| $\{p,\cdot,p\}$ |  |  | 0 | $2\hbar i\{p,\cdot,p\}$ | $-2\hbar i\{p,\cdot,p\}$ |
| $\{p,\cdot,q\}$ |  |  |  | 0 | $-2\hbar i\{p,\cdot,q\}$ $-2\hbar i\{q,\cdot,p\}$ |
| $\{q,\cdot,p\}$ |  |  |  |  | 0 |

Table (1)

For example, $[\{p,\cdot,q\},\{q,\cdot,p\}]=-2\hbar i(\{p,\cdot,q\}+\{q,\cdot,p\})$. Because the algebra is closed, we are prompted to consider the following ansatz (cf. Ref. [25]):

$$\rho(t) = e^{i(4\hbar)^{-1}w_4(\{q,\cdot,p\}-\{p,\cdot,q\})}e^{w_3(\{q,\cdot,p\}+\{p,\cdot,q\})}e^{-w_2\{p,\cdot,p\}}e^{-w_1\{q,\cdot,q\}}\mathcal{U}_{rev}\rho(0) \qquad (19)$$

where $\dfrac{d\mathcal{U}_{rev}}{dt} = \mathcal{L}_s \mathcal{U}_{rev}$ with $\mathcal{L}_s = \dfrac{1}{\hbar i}[H_s,\cdot]$. If we wish, by using the identity [37]

$$e^{r_1 A}e^{r_2 B} = \exp\left(r_1 A + \frac{r_1 r_2}{1-e^{-r_1}}B\right), \qquad (20)$$

where $r_1$ and $r_2$ are scalars and the two operators $A$ and $B$ satisfy $[A,B]=B$,

and some of the commutation relations in Table (1), we can combine all of the exponentials as follows:

$$\rho(t) = \exp\left[-\frac{w_4}{e^{w_4}-1}\left(w_1\{q,\cdot,q\}+w_2\{p,\cdot,p\}-\left(w_3-i\frac{e^{w_4}-1}{4\hbar}\right)\{p,\cdot,q\}-\left(w_3+i\frac{e^{w_4}-1}{4\hbar}\right)\{q,\cdot,p\}\right)\right]\mathcal{U}_{rev}\rho(0) \qquad (21)$$

The Wei-Norman method [38] can be used to relate the coefficients in the preceding exponent to the coefficients of the master equation. We find the following system of differential equations:

$$\frac{d}{dt}\begin{pmatrix}w_1\\w_2\\w_3\end{pmatrix} = 2\begin{pmatrix}-2b_{12} & 0 & -2b_{11}\\0 & 2b_{12} & 2b_{22}\\b_{22} & -b_{11} & 0\end{pmatrix}\begin{pmatrix}w_1\\w_2\\w_3\end{pmatrix} + e^{w_4}\begin{pmatrix}k_1\\k_2\\(k_3+k_4)/2\end{pmatrix} \qquad (22)$$

with

$$w_4 = 2\hbar i\int_0^t (k_3-k_4)dt', \qquad (23)$$

and initial conditions $w_1(0)=w_2(0)=w_3(0)=0$. For later use, we note the following identity that follows from the system (22):

$$w_1 w_2 - w_3^2 = \int_0^t e^{w_4}\left[k_1 w_2 + k_2 w_1 - (k_3+k_4)w_3\right]dt' \qquad (24)$$

The validity of this last equation can be demonstrated by differentiating both sides and using Eq. (22).

Using well known techniques, the solution of the system of differential equations (22) may be obtained readily:

$$\begin{pmatrix}w_1\\w_2\\w_3\end{pmatrix} = \int_0^t \Pi_w(t,s)e^{w_4(s)}\begin{pmatrix}k_1(s)\\k_2(s)\\[k_3(s)+k_4(s)]/2\end{pmatrix}ds \qquad (25)$$

where $\Pi_w(t,s)$ is the principal matrix solution of the associated homogeneous system

$$\frac{d}{dt}\begin{pmatrix} w_{1,h} \\ w_{2,h} \\ w_{3,h} \end{pmatrix} = 2 \begin{pmatrix} -2b_{12} & 0 & -2b_{11} \\ 0 & 2b_{12} & 2b_{22} \\ b_{22} & -b_{11} & 0 \end{pmatrix} \begin{pmatrix} w_{1,h} \\ w_{2,h} \\ w_{3,h} \end{pmatrix}. \tag{26}$$

The last homogeneous equation may be difficult to solve when the coefficients $b_{ij}$ depend on time. We therefore provide a technique that reduces the problem of solving Eq. (26) to one of solving one-dimensional harmonic oscillator equations with time dependent coefficients. To wit, as can be confirmed by first differentiating out the terms and then using Eq. (26) and $w_{1,h} w_{2,h} - w_{3,h}^2 = 0$ (cf. Eq. (24)), we find that

$$\frac{d}{dt}\left(\frac{d}{dt} \ln w_{1,h}\right) = \frac{\dot{b}_{11}}{b_{11}} \frac{d}{dt} \ln w_{1,h} - \frac{1}{2}\left(\frac{d}{dt} \ln w_{1,h}\right)^2 + 4\left(2b_{12}^2 + b_{12}\frac{\dot{b}_{11}}{b_{11}} - \dot{b}_{12} - 2b_{11}b_{22}\right). \tag{27}$$

This is a Riccati equation in the variable $\frac{d}{dt} \ln w_{1,h}$, and therefore calls out for the transformation $2\frac{d}{dt} \ln y_1 \equiv \frac{d}{dt} \ln w_{1,h}$. We get

$$\frac{d^2 y_1}{dt^2} = \frac{\dot{b}_{11}}{b_{11}} \frac{dy_1}{dt} + 2\left(2b_{12}^2 + b_{12}\frac{\dot{b}_{11}}{b_{11}} - \dot{b}_{12} - 2b_{11}b_{22}\right) y_1, \tag{28}$$

which is the equation of harmonic oscillator with time-dependent friction and frequency. Likewise, we find

$$\frac{d}{dt}\left(\frac{d}{dt} \ln w_{2,h}\right) = \frac{\dot{b}_{22}}{b_{22}} \frac{d}{dt} \ln w_{2,h} - \frac{1}{2}\left(\frac{d}{dt} \ln w_{2,h}\right)^2 + 4\left(2b_{12}^2 - b_{12}\frac{\dot{b}_{22}}{b_{22}} + \dot{b}_{12} - 2b_{11}b_{22}\right) \tag{29}$$

The transformation $2\frac{d}{dt} \ln y_2 \equiv \frac{d}{dt} \ln w_{2,h}$ leads to

$$\frac{d^2 y_2}{dt^2} = \frac{\dot{b}_{22}}{b_{22}} \frac{dy_2}{dt} + 2\left(2b_{12}^2 - b_{12}\frac{\dot{b}_{22}}{b_{22}} + \dot{b}_{12} - 2b_{11}b_{22}\right) y_2 \tag{30}$$

If we could find solutions of Eqs. (28) and (30), $(w_{1,h}, w_{2,h}, w_{3,h})$ could be built up therefrom, followed by the computation of $\Pi_w(t,s)$ and finally $(w_1, w_2, w_3)$. With the last trio in hand, the solution of Eq. (18) is given by Eq. (21).

*Example: Harmonic oscillator bilinearly coupled to a heat bath*

Consider a harmonic oscillator (system of interest) bilinearly coupled to an infinite number of other oscillators (reservoir or bath). The total Hamiltonian describing such a system of interest and reservoir is

$$H_T = \sum_{\nu=0}^{N}\left(\frac{p_\nu^2}{2m_\nu} + \frac{m_\nu \omega_\nu^2 q_\nu^2}{2}\right) + \sum_{n=1}^{N} \varepsilon_n q_0 q_n, \qquad (31)$$

where $(q_0, p_0)$ and $(q_1,...,q_N, p_1,...,p_N)$ are the canonical coordinates of the system of interest and reservoir, respectively. If the model represented by Hamiltonian (3) is replaced by one having continuous frequencies with Ullersma's [4] spectral strength function

$$f(\omega) = \frac{2}{\pi} \frac{\kappa \alpha^2 \omega^2}{\alpha^2 + \omega^2}, \qquad (32)$$

where $\alpha$ plays the role of a high frequency cut-off, $\kappa$ is a measure of the coupling strength between the system of interest and the reservoir, and $\omega$ are the frequencies of the reservoir oscillators, and if a factorized initial state is assumed (with the bath in thermal equilibrium), an exact master equation may be computed. This master equation was derived and solved in the Wigner representation by Haake and Reibold [5]. In our notation, this master equation reads:

$$\frac{d}{dt}\rho(t) = \frac{1}{\hbar i}\left[\frac{p^2}{2m} - \frac{1}{4}f_{pp}(t)(qp+pq) - \frac{m}{2}f_{pq}(t)q^2, \rho(t)\right] + \frac{1}{2\hbar^2}\left(d_{pq}(t) + i\frac{\hbar}{2}f_{pp}(t)\right)\{p,\rho(t),q\}$$
$$+ \frac{1}{2\hbar^2}\left(d_{pq}(t) - i\frac{\hbar}{2}f_{pp}(t)\right)\{q,\rho(t),p\} - \frac{m}{\hbar^2}d_{pp}(t)\{q,\rho(t),q\} \qquad (33)$$

where the coefficients, after correcting a couple of typographical errors, are provided in Ref. [5]:

$$f_{pq}(t) = -\left(\ddot{A}^2 - \dot{A}\dddot{A}\right)/R^2, \qquad (34)$$

$$f_{pp}(t) = -\left(A\dddot{A} - \dot{A}\ddot{A}\right)/R^2, \qquad (35)$$

$$d_{pp}(t) = \frac{1}{2}\dot{Y} - \frac{1}{2}f_{pq}\dot{X} - f_{pp}Y, \text{ and} \qquad (36)$$

$$d_{pq}(t) = -Y + \frac{1}{2}\ddot{X} - \frac{1}{2}f_{pp}\dot{X} - f_{pq}X, \qquad (37)$$

with $X(t) = \frac{\hbar}{2}\int_0^\infty d\omega \frac{f(\omega)}{\omega}\left|\int_0^t dt' e^{i\omega t'} A(t')\right|^2 \coth\left(\frac{\hbar\omega}{2kT}\right), \qquad (38)$

$$Y(t) = \frac{\hbar}{2}\int_0^\infty d\omega \frac{f(\omega)}{\omega}\left|\int_0^t dt' e^{i\omega t'} \dot{A}(t')\right|^2 \coth\left(\frac{\hbar\omega}{2kT}\right), \qquad (39)$$

$$A = \frac{2\Gamma[\exp((2\Gamma-\alpha)t) - \exp(-\Gamma t)\cos(\Omega t)] + \Omega^{-1}[(\alpha-2\Gamma)^2 + \Omega^2 - \Gamma^2]\exp(-\Gamma t)\sin(\Omega t)}{(\alpha - 3\Gamma)^2 + \Omega^2},$$

(40)

and

$$R = \sqrt{\dot{A}^2 - A\ddot{A}} \tag{41}$$

(we assume $\alpha \geq 3\Gamma$ to ensure that the last radicand is positive), such that

$$\Gamma = \frac{\kappa}{2}\frac{\alpha^2}{(\alpha-\Gamma)^2 + \Omega^2}, \quad \Omega^2 = \frac{\alpha\omega^2}{\alpha - 2\Gamma} - \Gamma^2, \text{ and } \omega^2 = \omega_0^2 - \alpha\kappa \text{ ($\omega$ must be non-negative}$$

for $H_T$ to have a minimum). For what follows, we note that $0 < R^2 \leq 1$, provided $\alpha \geq 3\Gamma$, and $R(0) = 1$.

Equation (33) is of the form of Eq. (18) and therefore can be solved using the foregoing method after plugging the appropriate coefficients of Eq. (33) into the inhomogeneous component of Eq. (22).

Noting that

$$w_4 = -\int_0^t f_{pp} dt', \tag{42}$$

we can solve for $(w_1, w_2, w_3)$:

$$(w_1, w_2, w_3) = \frac{1}{2\hbar^2 R^2}\left(mY, \frac{X}{m}, \frac{\dot{X}}{2}\right). \tag{43}$$

Using Eq. (21), we find the following operator solution [25].

$$\rho(t) = \exp\left[\frac{\ln R^2}{2\hbar^2(1-R^2)}\left(\begin{array}{c} mY\{q,\cdot,q\} + \frac{1}{m}X\{p,\cdot,p\} \\ -\frac{1}{2}\left(\dot{X} - i\hbar(1-R^2)\right)\{p,\cdot,q\} - \frac{1}{2}\left(\dot{X} + i\hbar(1-R^2)\right)\{q,\cdot,p\} \end{array}\right)\right]$$
$$\times N(t)\tilde{M}(t)\rho(0)\tilde{M}(t)^\dagger N(t)^\dagger, \tag{44}$$

where the unitary operators $N$ and $\tilde{M}$ are characterized in Appendix 2.

As shown in Appendix 2, and defining the interaction picture operator

$$\rho^I(t) = \tilde{M}^\dagger(t) N^\dagger(t) \rho(t) N(t)\tilde{M}(t), \tag{45}$$

Eq. (44) can be put into a manifestly positive form. To wit,

$$\rho'(t) = \exp\left\{\left[-\frac{\ln R^2}{2} - \frac{1}{2}\ln\left(1 + R^{-2}\left(\frac{1}{\hbar}\left(XY - \frac{1}{4}\dot{X}^2\right)^{1/2} + \frac{1}{2}(1-R^2)\right)\right)\right]\{B,\cdot,B\}\right\}$$

$$\times \exp\left[-\frac{1}{2}\ln\left(1 + R^{-2}\left(\frac{1}{\hbar}\left(XY - \frac{1}{4}\dot{X}^2\right)^{1/2} + \frac{1}{2}(1-R^2)\right)\right)\{B^\dagger,\cdot,B^\dagger\}\right]\rho(0) \quad (46)$$

where $B(t)$, characterized in Appendix 2, satisfies $[B(t), B^\dagger(t)] = 1$. Since $0 < R^2(t) \leq 1$, the factors in front of $\{B^\dagger,\cdot,B^\dagger\}$ and $\{B,\cdot,B\}$ are non-positive. With the use of relations (16) and (17), we immediately have a solution that is manifestly positive.

Like Eq. (33), exact master equations derived from other models typically have complicated coefficients. It would be eminently desirable to approximate the coefficients in equations like (33). However, as soon as we introduce an approximation, the equation is no longer exact and all bets are off as regards positivity. In the next section, we formulate conditions for the $k$'s that ensure positivity, thereby providing guidance for the types of approximations that can be introduced.

*Positivity conditions*

As the examples in Ref. [5] attest, exact non-autonomous master equations need not be of Lindblad form with time-dependent coefficients. Instead of expressions for $d\rho/dt$, one can examine one or more exponents of the propagator to probe positivity, as has been done for Brownian systems [25]. In this section, we relate coefficients in these exponents to coefficients in the master equation, thereby providing several sets of relations involving the $k$'s that are sufficient to ensure positivity.

In Ref. [26], a direct corollary of Lindblad's work [19] was presented [23] which when applied to Eq. (21) states that if $\rho(0)$ is an allowable initial state, then the density operator $\rho(t)$ is positive for any $t \geq 0$ at which $w_1(t) \geq 0$, $w_2(t) \geq 0$, and

$$w_1(t)w_2(t) - w_3^2(t) - \left(\frac{e^{w_4}-1}{4\hbar}\right)^2 \geq 0. \quad (47)$$

It is convenient to introduce variables

$$(u_1, u_2, u_3, u_4) \equiv \frac{\hbar}{2}\left(\eta^{-1}w_1 + \eta w_2, \eta^{-1}w_1 - \eta w_2, 2w_3, \frac{1}{2\hbar}(e^{w_4}-1)\right) \quad (48)$$

where $\eta$ is any positive constant introduced to make the $u_j$ dimensionless. Then inequality (47) is equivalent to

$$u_\mu u^\mu \geq 0, \quad (49)$$

where generally $v_\mu v^\mu \equiv v_1^2 - v_2^2 - v_3^2 - v_4^2$. The notation is warranted because, as shown in Appendix 3, under a metaplectic transformation, the variables $(u_1, u_2, u_3, u_4)$ transform to $(u'_1, u'_2, u'_3, u'_4)$ but $u_\mu u^\mu$ remains invariant:

$$u'_\mu u'^\mu = u_\mu u^\mu. \tag{50}$$

In terms of the variables $(u_1, u_2, u_3)$, the system of differential equations (22) becomes

$$\frac{d}{dt}\begin{pmatrix} u_1 \\ u_2 \\ u_3 \end{pmatrix} = 2\begin{pmatrix} 0 & -2b_{12} & -\eta^{-1}b_{11} + \eta b_{22} \\ -2b_{12} & 0 & -\eta^{-1}b_{11} - \eta b_{22} \\ -\eta^{-1}b_{11} + \eta b_{22} & \eta^{-1}b_{11} + \eta b_{22} & 0 \end{pmatrix}\begin{pmatrix} u_1 \\ u_2 \\ u_3 \end{pmatrix} + \frac{\hbar}{2}e^{w_4}\begin{pmatrix} \eta^{-1}k_1 + \eta k_2 \\ \eta^{-1}k_1 - \eta k_2 \\ k_3 + k_4 \end{pmatrix} \tag{51}$$

with initial conditions $u_1(0) = u_2(0) = u_3(0) = 0$, and where, again, $w_4 = 2\hbar i \int_0^t (k_3 - k_4)dt'$. The solution of the system (51) is

$$\begin{pmatrix} u_1 \\ u_2 \\ u_3 \end{pmatrix} = \frac{\hbar}{2}\int_0^t \Pi_u(t,s)\exp\left(2i\hbar\int_0^s (k_3(t') - k_4(t'))dt'\right)\begin{pmatrix} \eta^{-1}k_1(s) + \eta k_2(s) \\ \eta^{-1}k_1(s) - \eta k_2(s) \\ k_3(s) + k_4(s) \end{pmatrix} ds \tag{52}$$

and

$$u_4 = \frac{1}{4}\left[\exp\left(2\hbar i \int_0^t (k_3(t') - k_4(t'))dt'\right) - 1\right] \tag{53}$$

where $\Pi_u(t,s)$ is the principal matrix solution of the homogeneous system associated with Eq. (51).

We have thus formulated in terms of the $k$'s one set of sufficient conditions for Eq. (18) to be positive: provided
$u_1(t) \geq |u_2(t)|$ and the initial state is allowable, $\tag{54}$
then a sufficient condition for $\rho(t)$ to be positive is that relation (49) hold at time $t$. If $\rho(t)$ is to be positive for all $t \geq 0$, then it is sufficient that (49) and (54) hold for all $t \geq 0$. We emphasize that although it appears that relation (49) is given in terms of the $u$'s instead of the $k$'s, we can use expressions (52) and (53) to rewrite relation (49) in terms of the $k$'s.

For computational purposes, we can do better, however. Introducing
$$(h_1, h_2, h_3, h_4) \equiv \frac{1}{\hbar}\left(\eta^{-1}k_1 + \eta k_2, \eta^{-1}k_1 - \eta k_2, k_3 + k_4, i(k_3 - k_4)\right) \tag{55}$$

and using Eqs. (24), (47) and (48), we find that

$$\int_0^t u_\mu h^\mu \exp\left(2\hbar^2 \int_0^{t'} h_4 dt''\right) dt' \geq 0 \tag{56}$$

and conditions (54) are sufficient to ensure that $\rho(t)$ is positive.

Finally, since the exponent in relation (56) is real, we also find that $\rho(t)$ is positive at all times if conditions (54) hold and

$$u_\mu h^\mu \geq 0 \tag{57}$$

at all times. The quantity $u_\mu h^\mu$ is invariant under a metaplectic transformation since $u_\mu h^\mu = \hbar^{-2}(1+4u_4)^{-1} d(u_\mu u^\mu)/dt$ and $u_4 = u'_4$. Compared to relation (49), the relation (57) is probably easier to work with since it is linear in the $u$'s. With the $u_\mu$ given by Eqs. (52) and (53), and the $h_\mu$ given by Eq. (55), these are sufficient conditions for the $k$'s of Eq. (18) to ensure positivity [39].

*Discussion*

Notwithstanding the cautionary comments above, the standard QBE does have a role to play in describing Brownian motion. An accurate characterization of Brownian motion involves several time scales. One scale involves $1/\alpha$ where $\alpha$ is a high-frequency cut-off of the bath. Because the $\alpha \to \infty$ limit is not uniform in time [5], a boundary layer arises at times on the order of $1/\alpha$ that separates an inner limit and an outer limit. In the inner limit, an inner solution describes the rapid entanglement of the system of interest and the bath and is the key to preserving positivity [25]. In the outer limit, high-temperature and white-noise approximations can be invoked that lead to the standard QBE. In Ref. [25], a patch was constructed consisting of an inner propagator followed by an outer propagator, which is given by the standard QBE. The inner propagator decreases the domain of density operators that is subjected to the outer propagator. On this smaller domain, it was shown that the standard QBE does preserve positivity in an appropriate high-temperature regime [40] (see also [42] in connection with a spin system).

A related notion is that of initial slips [5,12]. In this approach, an outer solution is in effect propagated backwards to the initial time to identify effective initial data. However, one drawback of this approach is that the effective data need not be positive, and therefore propagation of the effective data using an outer propagator is not accurate in the inner limit. Moreover, it is not clear whether all effective initial data become positive in the outer limit, though the techniques developed here could be used to answer this question.

While autonomous QBE's have a role to play, with best results achieved by decreasing the domain of initial conditions (through patching or with the use of effective initial data), it appears that autonomous equations yielding completely positive evolution do not enter, at least not for systems that are reduced from an underlying total Hamiltonian when the coupling is bilinear in position and the initial total state is uncorrelated. For such systems, autonomous, completely positive QBEs cannot describe evolution initially. Moreover, it seems that if one wishes to describe the outer limit with an autonomous

QBE, the correct form is that of Eq. (1), or variants thereof [5,43], without a $\{p,\cdot,p\}$ term.

It should be emphasized that these comments apply to a master equation obtained from Hamiltonian systems in the short time approximation, which leads to the QBE, not to a master equation obtained in the secular approximation, which leads to the quantum optical master equation [44]. In the former case, the course graining time is much smaller than natural system periods; in the latter, the course graining time is much larger than the natural system periods. From a Brownian equation with no $\{p,\cdot,p\}$ term, it is possible to derive a quantum optical master equation and pick up such a term [21]. However, for the quintessential Brownian particle, the free particle, which has an "infinite natural period," there is no corresponding quantum optical master equation.

For Hamiltonian formulations, better approaches for studying quantum Brownian motion would involve finding uniform approximations for the coefficients in the associated exact master equation that incorporate the inner and outer limits, and that transition seamlessly from one region to the other. It is for this approach that the work herein is helpful. The positivity conditions provided above can guide us in choosing uniform approximations for these coefficients that at once describe the inner and outer limits, and that preserve positivity.

*Appendix 1*

The theorem in this appendix provides a necessary condition for positivity. The proof is constructive, furnishing an initial state that evolves into a non-positive state under certain conditions. Although the theorem is presented in a different manner than the one appearing in Ref. [26], it more or less contains the same substance. To prove the theorem, we first need the following Lemma that yields some expectation values. These values were presented in Ref. [26], but space considerations prevented a presentation of their derivation.

<u>Lemma</u>

For $t \geq 0$, suppose
$$\rho(t) = \exp[-\eta(t)\{q,\cdot,q\} - \xi(t)\{p,\cdot,p\} + \zeta(t)\{q,\cdot,p\} + \zeta^*(t)\{p,\cdot,q\}]\sigma(t) \tag{58}$$
where $\eta(t) \geq 0$ and $\xi(t) \geq 0$ are real continuous functions, $\zeta(t)$ is a complex function with continuous real and imaginary components and $\sigma(t)$ is an allowable density operator. Suppose further that
$$\mathrm{Re}^2\,\zeta(t) \leq \eta(t)\xi(t). \tag{59}$$
Then
$$\langle q^2 \rangle(t) \equiv Tr\rho(t)q^2$$
$$= e^{-4\hbar\,\mathrm{Im}\,\zeta(t)}Tr\sigma(t)q^2 + \frac{\hbar[1-\exp(-4\hbar\,\mathrm{Im}\,\zeta(t))]}{2\,\mathrm{Im}\,\zeta(t)}\xi(t), \tag{60}$$
$$\langle p^2 \rangle(t) = e^{-4\hbar\,\mathrm{Im}\,\zeta(t)}Tr\sigma(t)p^2 + \frac{\hbar[1-\exp(-4\hbar\,\mathrm{Im}\,\zeta(t))]}{2\,\mathrm{Im}\,\zeta(t)}\eta(t) \tag{61}$$
and
$$\langle qp + pq \rangle(t) = e^{-4\hbar\,\mathrm{Im}\,\zeta(t)}Tr\sigma(t)(qp+pq) + \frac{\hbar[1-\exp(-4\hbar\,\mathrm{Im}\,\zeta(t))]}{\mathrm{Im}\,\zeta(t)}\mathrm{Re}\,\zeta(t). \tag{62}$$

<u>Proof</u>

We shall only prove Eq. (60), as the proofs for the other two expectation values are similar.

We have, suppressing the time-dependence,
$$\rho = \exp[i\,\mathrm{Im}\,\zeta(\{q,\cdot,p\} - \{p,\cdot,q\}) - \eta\{q,\cdot,q\} - \xi\{p,\cdot,p\} + \mathrm{Re}\,\zeta(\{q,\cdot,p\} + \{p,\cdot,q\})]\sigma \tag{63}$$
To factor out the dissipation, we use identity (20). Using Table 1, we compute that

$$\left[\frac{i\,\mathrm{Im}\,\zeta}{-4\hbar\,\mathrm{Im}\,\zeta}(\{q,\cdot,p\} - \{p,\cdot,q\}), -\eta\{q,\cdot,q\} - \xi\{p,\cdot,p\} + \mathrm{Re}\,\zeta(\{q,\cdot,p\} + \{p,\cdot,q\})\right]$$
$$= -\eta\{q,\cdot,q\} - \xi\{p,\cdot,p\} + \mathrm{Re}\,\zeta(\{q,\cdot,p\} + \{p,\cdot,q\}) \tag{64}$$

with $r_1 = -4\hbar \operatorname{Im}\zeta$ and $r_2 = \dfrac{e^{4\hbar \operatorname{Im}\zeta}-1}{4\hbar \operatorname{Im}\zeta}$, we can apply relation (20) to arrive at

$$\rho = \exp[i\operatorname{Im}\zeta(\{q,\cdot,p\}-\{p,\cdot,q\})]\exp\left[-\frac{e^{4\hbar \operatorname{Im}\zeta}-1}{4\hbar \operatorname{Im}\zeta}(\eta\{q,\cdot,q\}+\xi\{p,\cdot,p\}-\operatorname{Re}\zeta(\{q,\cdot,p\}+\{p,\cdot,q\}))\right]\sigma \tag{65}$$

Now we seek to write

$$\eta\{q,\cdot,q\}+\xi\{p,\cdot,p\}-\operatorname{Re}\zeta(\{q,\cdot,p\}+\{p,\cdot,q\}) = \frac{\varpi}{\hbar}\lambda_+\{Q,\cdot,Q\}+\frac{\lambda_-}{\varpi\hbar}\{P,\cdot,P\} \tag{66}$$

where $\varpi$, introduced to get the dimensions right, is an arbitrary positive parameter having dimensions of mass/time (e.g., for a harmonic oscillator, we can choose $\varpi = m\omega$), and where
$\lambda_+ \geq 0$, $\lambda_- \geq 0$, $[Q,P] = \hbar i$ and

$$\begin{pmatrix} Q \\ P \end{pmatrix} = \begin{pmatrix} D & \dfrac{F}{\varpi} \\ \varpi C & E \end{pmatrix}\begin{pmatrix} q \\ p \end{pmatrix}, \tag{67}$$

the variables $(\lambda_-,\lambda_+,C,D,E,F)$ being dimensionless. These relations result in the following constraints

$$\eta = \frac{\varpi}{\hbar}(\lambda_+ D^2 + \lambda_- C^2), \tag{68}$$

$$\xi = \frac{1}{\varpi\hbar}(\lambda_+ F^2 + \lambda_- E^2), \tag{69}$$

$$\operatorname{Re}\zeta = -\frac{1}{\hbar}(\lambda_+ DF + \lambda_- CE) \tag{70}$$

and
$$DE - CF = 1. \tag{71}$$
Note, too, that the Cauchy-Schwarz inequality imposes the inequality
$$\eta\xi \geq \operatorname{Re}^2\zeta. \tag{72}$$
We have four equations (68)-(71) and six variables $(\lambda_-,\lambda_+,C,D,E,F)$; therefore, two variables are independent. Thus, there is no unique way to achieve the decomposition (66), but this will be of no consequence in what follows.

Noting that $[\{Q,\cdot,Q\},\{P,\cdot,P\}] = 0$, we have

$$\rho = \exp[i\operatorname{Im}\zeta(\{q,\cdot,p\}-\{p,\cdot,q\})]\exp\left[-\frac{e^{4\hbar \operatorname{Im}\zeta}-1}{4\hbar \operatorname{Im}\zeta}\frac{\varpi\lambda_+}{\hbar}\{Q,\cdot,Q\}\right]\exp\left[-\frac{e^{4\hbar \operatorname{Im}\zeta}-1}{4\hbar \operatorname{Im}\zeta}\frac{\lambda_-}{\varpi\hbar}\{P,\cdot,P\}\right]\sigma \tag{73}$$

In this form, we can readily compute the expectation values.

$$\langle q^2\rangle = Tr q^2 \exp[i\operatorname{Im}\zeta(\{q,\cdot,p\}-\{p,\cdot,q\})]\exp\left[-\frac{e^{4\hbar \operatorname{Im}\zeta}-1}{4\hbar \operatorname{Im}\zeta}\frac{\varpi\lambda_+}{\hbar}\{Q,\cdot,Q\}\right]\exp\left[-\frac{e^{4\hbar \operatorname{Im}\zeta}-1}{4\hbar \operatorname{Im}\zeta}\frac{\lambda_-}{\varpi\hbar}\{P,\cdot,P\}\right]\sigma \tag{74}$$

Now consider the operator $i(\{q,\cdot,p\}-\{p,\cdot,q\})$. Its dual is $4\hbar - i(\{q,\cdot,p\}-\{p,\cdot,q\})$. (This means $TrAi(\{q,\cdot,p\}-\{p,\cdot,q\})\rho = Tr\rho[4\hbar - i(\{q,\cdot,p\}-\{p,\cdot,q\})]A$.) Hence,

$$\langle q^2 \rangle = e^{4\hbar \operatorname{Im}\zeta} Tr\left[\exp[-i\operatorname{Im}\zeta(\{q,\cdot,p\}-\{p,\cdot,q\})]q^2\right]\exp\left[-\frac{e^{4\hbar \operatorname{Im}\zeta}-1}{4\hbar \operatorname{Im}\zeta}\frac{\varpi \lambda_+}{\hbar}\{Q,\cdot,Q\}\right]\exp\left[-\frac{e^{4\hbar \operatorname{Im}\zeta}-1}{4\hbar \operatorname{Im}\zeta}\frac{\lambda_-}{\varpi \hbar}\{P,\cdot,P\}\right]\sigma$$

(75)

We further note that if $f(q,p)$ is a symmetrized (Weyl-ordered) function of the operators $q$ and $p$ and $\tau$ is a real number then

$$\exp[\tau(\{q,\cdot,p\}-\{p,\cdot,q\})]f(q,p) = e^{4\hbar\tau}f(qe^{2\hbar\tau}, pe^{2\hbar\tau}).$$

(76)

Hence,

$$\langle q^2 \rangle = e^{-4\hbar \operatorname{Im}\zeta} Tr q^2 \exp\left[-\frac{e^{4\hbar \operatorname{Im}\zeta}-1}{4\hbar \operatorname{Im}\zeta}\frac{\varpi \lambda_+}{\hbar}\{Q,\cdot,Q\}\right]\exp\left[-\frac{e^{4\hbar \operatorname{Im}\zeta}-1}{4\hbar \operatorname{Im}\zeta}\frac{\lambda_-}{\varpi \hbar}\{P,\cdot,P\}\right]\sigma$$

$$= e^{-4\hbar \operatorname{Im}\zeta} Tr\left(EQ - \frac{1}{\varpi}FP\right)^2 \exp\left[-\frac{e^{4\hbar \operatorname{Im}\zeta}-1}{4\hbar \operatorname{Im}\zeta}\frac{\varpi \lambda_+}{\hbar}\{Q,\cdot,Q\}\right]\exp\left[-\frac{e^{4\hbar \operatorname{Im}\zeta}-1}{4\hbar \operatorname{Im}\zeta}\frac{\lambda_-}{\varpi \hbar}\{P,\cdot,P\}\right]\sigma$$

(77)

where we used the inverse of Eq. (67),

$$\begin{pmatrix} q \\ p \end{pmatrix} = \begin{pmatrix} E & -\frac{1}{\varpi}F \\ -\varpi C & D \end{pmatrix}\begin{pmatrix} Q \\ P \end{pmatrix},$$

(78)

to write down the last equation for $\langle q^2 \rangle$. But for any scalar $\tau > 0$ and any self-adjoint operator $A$, we have [45,25]

$$e^{-\tau\{A,\cdot,A\}}\rho = \frac{1}{2\pi}\sqrt{\frac{\pi}{\tau}}\int_{-\infty}^{\infty} du \exp\left(-\frac{u^2}{4\tau}\right)e^{-iuA}\rho e^{iuA}.$$

(79)

We can use this result twice, first for the exponential factor containing $\{Q,\cdot,Q\}$, and then for the factor containing $\{P,\cdot,P\}$. Using Eq. (67) to revert back to original variables $q$ and $p$, and then using relations (69) and (71), we finally obtain

$$\langle q^2 \rangle(t) = e^{-4\hbar \operatorname{Im}\zeta(t)}Tr\sigma(t)q^2 + \frac{1-e^{-4\hbar \operatorname{Im}\zeta(t)}}{\operatorname{Im}\zeta(t)}\frac{\hbar\xi(t)}{2}.$$

The foregoing Lemma is needed to prove the following theorem.

<u>Theorem</u>

For $t \geq 0$, suppose

$$\rho(t) = \exp\left[-\eta(t)\{q,\cdot,q\} - \xi(t)\{p,\cdot,p\} + \zeta(t)\{q,\cdot,p\} + \zeta^*(t)\{p,\cdot,q\}\right]U(t)\rho(0)U^\dagger(t) \quad (80)$$

where $\eta(t) \geq 0$ and $\xi(t) \geq 0$ are real continuous functions, $\zeta(t)$ is a complex function with continuous real and imaginary components, and $U(t)$ is a unitary operator with $U(0)=1$.

We assume $\eta(0) = \xi(0) = \zeta(0) = 0$ and that $\rho(0)$ is an allowable initial density operator. Suppose further that $\text{Im}\,\zeta(t') > 0$, $\eta(t') > 0$ and
$$0 \leq \eta(t')\xi(t') - \text{Re}^2\,\zeta(t') < \text{Im}^2\,\zeta(t') \tag{81}$$
at some time $t' > 0$. Then there exists an allowable density operator $\sigma$ such that $\rho(t')$ is non-positive when $\rho(0) = \sigma$.

Proof

First, we construct the allowable density operator $\sigma = U^\dagger(t')\chi U(t')$, where $\chi$ is the pure, allowable density operator corresponding to the Wigner function
$$W(p,q) = \frac{1}{\pi\hbar} \exp\left[-\frac{2}{\hbar^2}\left(\lambda_3 q^2 + \lambda_1 p^2 - \lambda_2 pq\right)\right], \tag{82}$$
such that
$$d_p \equiv (\hbar\,\text{Im}\,\zeta + \lambda_2\,\text{Re}\,\zeta)^2 - \eta\xi(\hbar^2 + \lambda_2^2) > 0, \tag{83}$$
$$\hbar\,\text{Im}\,\zeta + \lambda_2\,\text{Re}\,\zeta > 0, \tag{84}$$
where these last two and all other expressions in this proof are evaluated at time $t'$ unless otherwise indicated, and
$$\lambda_3 = \frac{\hbar^2 + \lambda_2^2}{4\lambda_1} \tag{85}$$
with $\lambda_1 > 0$ being specified below. Note that for this Wigner function, the expectation values of $p^2, q^2$ and $qp + pq$ are $\lambda_3$, $\lambda_1$ and $\lambda_2$, respectively.

Under the hypotheses of the theorem, let us first show that $\lambda_1$ and $\lambda_2$ with the foregoing restrictions exist. We will consider the cases $\text{Re}\,\zeta \neq 0$ and $\text{Re}\,\zeta = 0$ separately.

i) Assume first that $\text{Re}\,\zeta \neq 0 \Rightarrow \eta\xi > 0$. Then, relation (83) is equivalent to
$$d_p\,\text{Re}^2\,\zeta = (\text{Re}^2\,\zeta - \eta\xi)(\hbar\,\text{Im}\,\zeta + \lambda_2\,\text{Re}\,\zeta)^2 + 2\eta\xi\hbar(\text{Im}\,\zeta)(\hbar\,\text{Im}\,\zeta + \lambda_2\,\text{Re}\,\zeta) - \eta\xi\hbar^2|\zeta|^2 > 0. \tag{86}$$

If $\text{Re}^2\,\zeta - \eta\xi = 0$, then we choose a $\lambda_2$ such that $\hbar\,\text{Im}\,\zeta + \lambda_2\,\text{Re}\,\zeta > \dfrac{\hbar|\zeta|^2}{2\,\text{Im}\,\zeta}$, and this ensures relations (83) and (84) hold.

Now, suppose $\text{Re}^2\,\zeta - \eta\xi \neq 0$ and consider
$$(\text{Re}^2\,\zeta - \eta\xi)(\hbar\,\text{Im}\,\zeta + \lambda_2\,\text{Re}\,\zeta)^2 + 2\eta\xi\hbar(\text{Im}\,\zeta)(\hbar\,\text{Im}\,\zeta + \lambda_2\,\text{Re}\,\zeta) - \eta\xi\hbar^2|\zeta|^2 \tag{87}$$

as a quadratic function in $\hbar\,\text{Im}\,\zeta + \lambda_2\,\text{Re}\,\zeta$. Because the discriminant is positive, the quadratic has two real roots. Since $\text{Re}^2\,\zeta - \eta\xi < 0$, the quadratic is concave down, and moreover, because the largest root is positive, it is always possible to choose a $\hbar\,\text{Im}\,\zeta + \lambda_2\,\text{Re}\,\zeta > 0$ such that $d_p > 0$.

ii) Now assume $\text{Re}\,\zeta = 0 \Rightarrow 0 \leq \eta\xi < \text{Im}^2\,\zeta$. Then $d_p = (\hbar\,\text{Im}\,\zeta)^2 - \eta\xi(\hbar^2 + \lambda_2^2)$, and we see that if $\eta\xi = 0$ then $d_p > 0$ for any value of $\hbar\,\text{Im}\,\zeta + \lambda_2\,\text{Re}\,\zeta$. So assume $\eta\xi > 0$. Then $d_p = (\hbar\,\text{Im}\,\zeta)^2 - \eta\xi(\hbar^2 + \lambda_2^2) > 0$ if $\lambda_2^2 < \dfrac{\hbar^2}{\eta\xi}(\text{Im}^2\,\zeta - \eta\xi)$. Moreover, $\hbar\,\text{Im}\,\zeta + \lambda_2\,\text{Re}\,\zeta > 0$, since we are assuming $\text{Re}\,\zeta = 0$.

To summarize, under the hypotheses of the theorem, it is always possible to find $\lambda_2$ such that $d_p > 0$ and $\hbar\,\text{Im}\,\zeta + \lambda_2\,\text{Re}\,\zeta > 0$. Below we assume $\lambda_2$ is so chosen. We now outline how $\lambda_1 > 0$ is to be chosen.

Consider the quadratic form in $x$:
$$\eta x^2 - (\hbar\,\text{Im}\,\zeta + \lambda_2\,\text{Re}\,\zeta)x + \frac{1}{4}\xi(\hbar^2 + \lambda_2^2). \tag{88}$$
Its discriminant is $d_p > 0$.

Therefore, the quadratic has two distinct real roots:
$$s_{\pm} = \frac{1}{2\eta}\left(\hbar\,\text{Im}\,\zeta + \lambda_2\,\text{Re}\,\zeta \pm d_p^{1/2}\right). \tag{89}$$
Also, because $\hbar\,\text{Im}\,\zeta + \lambda_2\,\text{Re}\,\zeta > 0$, $\eta > 0$ and $\xi \geq 0$, the smaller root $s_-$ is non-negative and the larger root $s_+$ is positive. We choose
$$\lambda_1 = \frac{1}{2}(s_- + s_+) > 0. \tag{90}$$
Having chosen $\lambda_1$ and $\lambda_2$, we fix $\lambda_3$ according to Eq. (85).

Now note that
$$\eta\lambda_1^2 - (\hbar\,\text{Im}\,\zeta + \lambda_2\,\text{Re}\,\zeta)\lambda_1 + \frac{1}{4}\xi(\hbar^2 + \lambda_2^2) < 0. \tag{91}$$
Multiplying by $-2\hbar\,\dfrac{\exp(-4\hbar\,\text{Im}\,\zeta)}{\lambda_1}\left(\dfrac{1-\exp(-4\hbar\,\text{Im}\,\zeta)}{\text{Im}\,\zeta}\right)$ and using Eq. (85), we get
$$2\hbar\exp(-4\hbar\,\text{Im}\,\zeta)\left(\frac{1-\exp(-4\hbar\,\text{Im}\,\zeta)}{\text{Im}\,\zeta}\right)(\hbar\,\text{Im}\,\zeta + \lambda_2\,\text{Re}\,\zeta - \eta\lambda_1 - \xi\lambda_3) > 0. \tag{92}$$
Now introduce a parameter
$$\lambda = \frac{\hbar + w^{1/2}}{2\left[\lambda_3\exp(-4\hbar\,\text{Im}\,\zeta) + \dfrac{\hbar(1-\exp(-4\hbar\,\text{Im}\,\zeta))}{2\,\text{Im}\,\zeta}\eta\right]}, \tag{93}$$
where $w$ is defined to be the left-hand side of inequality (92). By virtue of inequality (92), $\lambda$ is real, and by construction, $\lambda$ is a root that satisfies

$$\left[\lambda_3 \exp(-4\hbar \operatorname{Im}\zeta) + \frac{\hbar(1-\exp(-4\hbar \operatorname{Im}\zeta))}{2\operatorname{Im}\zeta}\eta\right]\lambda^2 - \hbar\lambda + \frac{\hbar^2 - w}{4\left[\lambda_3 \exp(-4\hbar \operatorname{Im}\zeta) + \frac{\hbar(1-\exp(-4\hbar \operatorname{Im}\zeta))}{2\operatorname{Im}\zeta}\eta\right]} = 0$$

(94)

In another vein, letting $\beta$ be a real parameter, let us expand

$$I(\beta,\lambda;t') \equiv Tr\,[q + (\beta + i\lambda(t'))p]\rho(t')[q + (\beta - i\lambda(t'))p]$$
$$= \beta^2\langle p^2\rangle + \beta\langle qp + pq\rangle + \lambda^2\langle p^2\rangle - \lambda\hbar + \langle q^2\rangle. \qquad (95)$$

But for evolution governed by Eq. (80), the results (60)-(62) of the previous Lemma apply. Noting that at time $t'$ we have

$$\lambda_1 = Tr q^2 U(t')\sigma U^\dagger(t'), \qquad (96)$$
$$\lambda_2 = Tr(qp + pq)U(t')\sigma U^\dagger(t'), \qquad (97)$$

and

$$\lambda_3 = Tr p^2 U(t')\sigma U^\dagger(t') \qquad (98)$$

we obtain

$$I(\beta,\lambda;t')$$
$$= \beta^2\left[e^{-4\hbar\operatorname{Im}\zeta}\lambda_3 + \frac{\hbar[1-\exp(-4\hbar\operatorname{Im}\zeta)]}{2\operatorname{Im}\zeta}\eta\right] + \beta\left[e^{-4\hbar\operatorname{Im}\zeta}\lambda_2 + \frac{\hbar[1-\exp(-4\hbar\operatorname{Im}\zeta)]}{\operatorname{Im}\zeta}\operatorname{Re}\zeta\right]$$
$$+ \lambda^2\left[e^{-4\hbar\operatorname{Im}\zeta}\lambda_3 + \frac{\hbar[1-\exp(-4\hbar\operatorname{Im}\zeta)]}{2\operatorname{Im}\zeta}\eta\right] - \lambda\hbar + e^{-4\hbar\operatorname{Im}\zeta}\lambda_1 + \frac{\hbar[1-\exp(-4\hbar\operatorname{Im}\zeta)]}{2\operatorname{Im}\zeta}\xi.$$

(99)

Considering the right hand side of Eq. (99) as a quadratic function in $\beta$, we conclude that $I(\lambda,\bar\beta;t') < 0$ provided the discriminant satisfies

$$\left[e^{-4\hbar\operatorname{Im}\zeta}\lambda_2 + \frac{\hbar[1-\exp(-4\hbar\operatorname{Im}\zeta)]}{\operatorname{Im}\zeta}\operatorname{Re}\zeta\right]^2$$
$$- 4\left[e^{-4\hbar\operatorname{Im}\zeta}\lambda_3 + \frac{\hbar[1-\exp(-4\hbar\operatorname{Im}\zeta)]}{2\operatorname{Im}\zeta}\eta\right]\left[\begin{array}{c}e^{-4\hbar\operatorname{Im}\zeta}\lambda_3\lambda^2 + \frac{\hbar[1-\exp(-4\hbar\operatorname{Im}\zeta)]}{2\operatorname{Im}\zeta}\eta\lambda^2 \\ -\lambda\hbar + e^{-4\hbar\operatorname{Im}\zeta}\lambda_1 + \frac{\hbar[1-\exp(-4\hbar\operatorname{Im}\zeta)]}{2\operatorname{Im}\zeta}\xi\end{array}\right] > 0$$

(100)

and provided $\bar\beta$ lies between the two roots of the right hand side of Eq. (99), which are both real when inequality (100) is imposed. It is straightforward but tedious to show that when Eqs. (85) and (94) hold, inequality (100) is equivalent to the simpler inequality $\eta(t')\xi(t') < |\zeta(t')|^2$, which is part of assumption (81). We have thus shown that with the foregoing assumptions, $I(\lambda, \beta; t')$ is negative when $\lambda$ and $\beta$ are given by expression (93) and $\bar{\beta}$, respectively. This implies $\rho(t')$ is non-positive, which completes the proof.

*Appendix 2*

Here we show that the exact solution (44),

$$\rho(t) = \exp\left[\frac{\ln R^2}{2\hbar^2(1-R^2)}\left(\begin{array}{c} mY\{q,\cdot,q\} + \frac{1}{m}X\{p,\cdot,p\} \\ -\frac{1}{2}\left(\dot{X} - i\hbar(1-R^2)\right)\{p,\cdot,q\} - \frac{1}{2}\left(\dot{X} + i\hbar(1-R^2)\right)\{q,\cdot,p\} \end{array}\right)\right]$$
$$\times N(t)\tilde{M}(t)\rho(0)\tilde{M}(t)^\dagger N(t)^\dagger,$$

can be re-written in manifestly positive form as

$$\rho'(t) = \exp\left\{\left[-\frac{\ln R^2}{2} - \frac{1}{2}\ln\left(1 + R^{-2}\left(\frac{1}{\hbar}\left(XY - \frac{1}{4}\dot{X}^2\right)^{1/2} + \frac{1}{2}(1-R^2)\right)\right)\right]\{B,\cdot,B\}\right\}$$
$$\times \exp\left[-\frac{1}{2}\ln\left(1 + R^{-2}\left(\frac{1}{\hbar}\left(XY - \frac{1}{4}\dot{X}^2\right)^{1/2} + \frac{1}{2}(1-R^2)\right)\right)\{B^\dagger,\cdot,B^\dagger\}\right]\rho(0)$$

where $\rho'(t) = \tilde{M}^\dagger(t) N^\dagger(t) \rho(t) N(t) \tilde{M}(t)$ and where time-dependent $B$ satisfies $[B(t), B^\dagger(t)] = 1$.

To this end, we first note the following properties of the unitary operators $M$ and $N$ [25]:

$$N(t)^\dagger q\, N(t) = Eq - \frac{F}{m\omega}p \qquad (101)$$

$$N(t)^\dagger p\, N(t) = -m\omega Cq + Dp \qquad (102)$$

where

$$\begin{pmatrix} C & D \\ E & F \end{pmatrix} = \frac{1}{\sqrt{\left(\frac{\dot{X}}{\hbar}\right)^2 + \left(\frac{Y}{\hbar\omega} - \frac{\omega}{\hbar}X + \sqrt{\left(\frac{\dot{X}}{\hbar}\right)^2 + \left(\frac{Y}{\hbar\omega} - \frac{\omega}{\hbar}X\right)^2}\right)^2}}$$
$$\times \begin{pmatrix} \frac{\dot{X}}{\hbar} & \frac{Y}{\hbar\omega} - \frac{\omega}{\hbar}X + \sqrt{\left(\frac{\dot{X}}{\hbar}\right)^2 + \left(\frac{Y}{\hbar\omega} - \frac{\omega}{\hbar}X\right)^2} \\ \frac{Y}{\hbar\omega} - \frac{\omega}{\hbar}X + \sqrt{\left(\frac{\dot{X}}{\hbar}\right)^2 + \left(\frac{Y}{\hbar\omega} - \frac{\omega}{\hbar}X\right)^2} & -\frac{\dot{X}}{\hbar} \end{pmatrix}$$

(103)

and

$$\tilde{M}^{\dagger}(t)q\tilde{M}(t)=\frac{1}{R}\left[\left(\dot{A}D+\ddot{A}F/\omega\right)q+\frac{1}{m}\left(AD+\dot{A}F/\omega\right)p\right] \quad (104)$$

$$\tilde{M}^{\dagger}(t)p\tilde{M}(t)=\frac{1}{R}\left[m\left(\omega\dot{A}C+\ddot{A}E\right)q+\left(\omega AC+\dot{A}E\right)p\right] \quad (105)$$

where $R=\sqrt{\dot{A}^2-A\ddot{A}}$ (we assume $\alpha \geq 3\Gamma$ to ensure that the last radicand is positive). The combined action of $M$ and $N$ may be obtained from the preceding results:

$$\tilde{M}^{\dagger}(t)N^{\dagger}(t)qN(t)\tilde{M}(t)=\frac{1}{R}\left(\dot{A}q+\frac{1}{m}Ap\right) \quad (106)$$

$$\tilde{M}^{\dagger}(t)N^{\dagger}(t)pN(t)\tilde{M}(t)=\frac{1}{R}\left(m\ddot{A}q+\dot{A}p\right) \quad (107)$$

Using these relations, we can slide the unitary operators to the outside in expression (44) to obtain

$$\rho'(t)=\exp\left[\frac{\ln R^2}{2\hbar^2 R^2(1-R^2)}\begin{pmatrix} m\left(\dot{A}^2 Y+\ddot{A}^2 X-\dot{A}\ddot{A}\dot{X}\right)\{q,\cdot,q\} \\ +\left(A\ddot{A}X+A\dot{A}Y-\frac{1}{2}\left(\dot{A}^2+A\ddot{A}\right)\dot{X}-\frac{1}{2}i\hbar(1-R^2)R^2\right)\{q,\cdot,p\} \\ +\left(A\ddot{A}X+A\dot{A}Y-\frac{1}{2}\left(\dot{A}^2+A\ddot{A}\right)\dot{X}+\frac{1}{2}i\hbar(1-R^2)R^2\right)\{p,\cdot,q\} \\ +\frac{1}{m}\left(A^2 Y+\dot{A}^2 X-A\dot{A}\dot{X}\right)\{p,\cdot,p\} \end{pmatrix}\right]\rho(0)$$

$$(108)$$

The exponent is of the form
$$-a\{q,\cdot,q\}+c\{q,\cdot,p\}+c^*\{p,\cdot,q\}-b\{p,\cdot,p\}, \quad (109)$$

with $a$ and $b$ being real parameters, and $c$ being complex, and we wish to write this as

$$-a\{q,\cdot,q\}+c\{q,\cdot,p\}+c^*\{p,\cdot,q\}-b\{p,\cdot,p\}=-k_1\{B,\cdot,B\}-k_2\{B^{\dagger},\cdot,B^{\dagger}\}$$
$$\equiv -k_1\{rq+sp,\cdot,rq+sp\}-k_2\{tq+up,\cdot,tq+up\} \quad (110)$$

where $[B,B^{\dagger}]=1$, $k_1 \geq 0$ and $k_2 \geq 0$. Here, $r,s,t$ and $u$ are complex parameters with components $r=r_1+ir_2$, etc. We have ten variables $(r_1,r_2,s_1,s_2,t_1,t_2,u_1,u_2,k_1,k_2)$ that can depend on time, but they are not all independent. Since we must have $rq+sp=t^*q+u^*p$, we obtain the relations $r=t^*$ and $s=u^*$, which allows us to remove four dependent variables leaving us with six variables, say $(r_1,r_2,s_1,s_2,k_1,k_2)$, but these are still not independent. The last commutator gives us

$$ru - st = \frac{1}{\hbar i}. \tag{111}$$

We also have
$$a = k_1|r|^2 + k_2|t|^2 \geq 0 \tag{112}$$
$$b = k_1|s|^2 + k_2|u|^2 \geq 0 \tag{113}$$
and
$$c = -\left(k_1 r^* s + k_2 t^* u\right) \tag{114}$$
from which the Cauchy-Schwarz inequality yields
$$ab \geq |c|^2. \tag{115}$$

It is convenient to switch to polar coordinates

$$r = R' e^{i\theta} \tag{116}$$
and
$$s = S' e^{i\phi}. \tag{117}$$

Eqs. (111)-(114) give us

$$R' = \left(\frac{a}{k_1 + k_2}\right)^{1/2} \tag{118}$$

$$S' = \left(\frac{b}{k_1 + k_2}\right)^{1/2} \tag{119}$$

$$k_2 - k_1 = 2\hbar \,\mathrm{Im}(c) \tag{120}$$

$$\cos(\phi - \theta) = -\frac{\mathrm{Re}(c)}{R' S'(k_1 + k_2)} \tag{121}$$

and

$$\sin(\phi - \theta) = \frac{1}{2\hbar R' S'}, \tag{122}$$

whence

$$\tan(\phi - \theta) = -\frac{k_1 + k_2}{2\hbar \,\mathrm{Re}(c)}, \tag{123}$$

with $0 \leq \phi - \theta \leq \pi$. Moreover, Eqs. (121) and (122) yield
$$k_1 + k_2 = 2\hbar \left(ab - \mathrm{Re}^2(c)\right)^{1/2},$$
where we assume $ab - \mathrm{Re}^2(c) > 0$ for $t > 0$.

Summarizing, of the six variables $(R', S', \phi - \theta, \phi + \theta, k_2 + k_1, k_2 - k_1)$, all are fixed except for $\phi + \theta$.

With the foregoing in mind, let us rewrite Eq. (108) as

$$\rho^I(t) = \exp\left(-a\{q,\cdot,q\} + c\{q,\cdot,p\} + c^*\{p,\cdot,q\} - b\{p,\cdot,p\}\right)\rho(0)$$
$$= \exp(-k_1\{B,\cdot,B\} - k_2\{B^\dagger,\cdot,B^\dagger\})\rho(0) \tag{124}$$

where $a, b$ and $c$ can be picked off from the exponent of Eq. (108) and where there exists some freedom in how we can choose $B$ reflecting the one aforementioned independent angle variable that we have at our disposal. Using the commutation relation

$$[\{B,\cdot,B\},\{B^\dagger,\cdot,B^\dagger\}] = -2(\{B,\cdot,B\} + \{B^\dagger,\cdot,B^\dagger\})$$
$$\Rightarrow$$
$$[\{B,\cdot,B\},\{B,\cdot,B\} + \{B^\dagger,\cdot,B^\dagger\}] = -2(\{B,\cdot,B\} + \{B^\dagger,\cdot,B^\dagger\}),$$

which can be obtained from $[B, B^\dagger] = 1$, and using relation (20), we can separate the raising and lowering operators in Eq. (124):

$$\rho^I(t) = \exp\left[\left(k_2 - k_1 - \frac{1}{2}\ln\left(1 + k_2 \frac{e^{2(k_2 - k_1)} - 1}{k_2 - k_1}\right)\right)\{B,\cdot,B\}\right]$$
$$\times \exp\left[-\frac{1}{2}\ln\left(1 + k_2 \frac{e^{2(k_2 - k_1)} - 1}{k_2 - k_1}\right)\{B^\dagger,\cdot,B^\dagger\}\right]\rho(0)$$

Heeding Eq. (120) and $k_1 + k_2 = 2\hbar(ab - \text{Re}^2(c))^{1/2}$, which leaves us one independent variable to fix $B$, we obtain Eq. (46).

*Appendix 3*

Here we show that $u_\mu u^\mu$ is invariant under metaplectic transformations. The well-known metaplectic operators, $M$, are unitary operators that give rise to linear transformations of the canonical operators [46]:

$$M^\dagger \begin{pmatrix} q \\ p \end{pmatrix} M = \begin{pmatrix} A & B \\ C & D \end{pmatrix} \begin{pmatrix} q \\ p \end{pmatrix} \tag{125}$$

with
$$AD - BC = 1. \tag{126}$$

Now suppose we subject a propagator of density operators, $e^{L(t)}$, where $L(t)$ is the exponent of Eq. (21), to a metaplectic transformation:
$$e^{L'} \equiv M^\dagger [e^L (M \cdot M^\dagger)] M \tag{127}$$
We can compute this action explicitly:
$$e^L = \exp\left[-\frac{w_4}{e^{w_4}-1}\left(w_1\{q,\because,q\} + w_2\{p,\because,p\} - \left(w_3 - i\frac{e^{w_4}-1}{4\hbar}\right)\{p,\because,q\} - \left(w_3 + i\frac{e^{w_4}-1}{4\hbar}\right)\{q,\because,p\}\right)\right]$$

$\rightarrow$

$$e^{L'} = \exp\left[-\frac{w_4}{e^{w_4}-1}\left(w'_1\{q,\because,q\} + w'_2\{p,\because,p\} - \left(w'_3 - i\frac{e^{w'_4}-1}{4\hbar}\right)\{p,\because,q\} - \left(w'_3 + i\frac{e^{w'_4}-1}{4\hbar}\right)\{q,\because,p\}\right)\right]$$
$$\tag{128}$$

where
$$\begin{pmatrix} w'_1 \\ w'_2 \\ w'_3 \\ w'_4 \end{pmatrix} = \begin{pmatrix} A^2 w_1 + C^2 w_2 - 2AC w_3 \\ B^2 w_1 + D^2 w_2 - 2BD w_3 \\ -AB w_1 - CD w_2 + (AD+BC) w_3 \\ w_4 \end{pmatrix}. \tag{129}$$

This last expression is obtained by inserting the metaplectic operators in (127) into the exponent of $e^L$ and computing terms like $M^\dagger \{q, M \cdot M^\dagger, q\} M$, etc. After collecting terms, we arrive at Eq. (129). From this last result and Eq. (48), it is straightforward but tedious to show $u'_\mu u'^\mu = u_\mu u^\mu$.